\documentclass[a4paper]{jpconf}
\usepackage{graphicx}
\usepackage{hyperref}

\usepackage{amsmath}
\usepackage{cite}
\usepackage{mathrsfs}
\usepackage[percent]{overpic}

\def\bc{\begin{center}}
\def\ec{\end{center}}
\def\beq{\begin{equation}}
\def\eeq{\end{equation}}
\def\beq*{\begin{equation*}}
\def\eeq*{\end{equation*}}

\def\av#1{\langle #1 \rangle}
\def\avr#1#2{\langle {#1} \rangle^{}_{#2}}

\def\bc{\begin{center}}
\def\ec{\end{center}}
\def\beq{\begin{equation}}
\def\eeq{\end{equation}}

\def\av#1{\langle #1 \rangle}

\def\avr#1#2{\langle {#1} \rangle^{}_{#2}}

\def\brel{b^{rel}}

\begin{document}
\title{Forward-backward correlations between intensive observables}

\author{Vladimir Kovalenko, Vladimir Vechernin}
\address{Saint Petersburg State University, 7/9 Universitetskaya nab., St. Petersburg, 199034, Russia}

\ead{v.kovalenko@spbu.ru, v.vechernin@spbu.ru}

\begin{abstract}
We demonstrate that the investigations of the forward-backward correlations between intensive observables enable to obtain more clear signal about the initial stage of hadronic interaction, e.g. about the process of string fusion, compared to usual forward-backward multiplicity correlations. As an example, the correlation between mean-event transverse momenta of charged particles in separated rapidity intervals is considered. 
We performed calculations in the framework of dipole-based Monte Carlo string fusion model. We obtained the dependence of the correlation strength on the collision centrality for different initial energies and colliding systems. It is shown that the dependence reveals the decline of the correlation coefficient for most central  Pb-Pb collisions at LHC energy.
We compare the results both with the ones obtained in alternative models and with the ones obtained by us using various MC generators.
\end{abstract}

\section{Forward-Backward Rapidity Correlations}
%
%
%
%
%
%
%
%
%

The study of the correlations between observables in two separated rapidity windows
(the so-called long-range forward-backward correlations) has been proposed \cite{Amelin:1994mf}
as a signature of the string fusion and percolation phenomenon \cite{Braun:1992ss,Braun:1991dg},
which is one of the collectivity effects in ultrarelativistic heavy ion collisions.
Later it was realized \cite{Braun:2000hd, Alessandro:2006yt, Vechernin:2007zza, Vechernin:2007zz} that the investigations of the forward-backward correlations between intensive observables,
such e.g. as mean-event transverse momenta,
enable to obtain more clear signal about the initial stage of hadronic interaction,
including the process of string fusion, compared to usual forward-backward multiplicity correlations.
 As an example, the correlation between mean-event transverse momenta of charged particles in separated rapidity intervals can be considered.

Numerically, the long-range correlations are studied in terms of correlation functions and
correlation coefficients.
Correlation function is defined as the mean value of variable
B in the backward window as a function of another variable F in the forward rapidity
window:
$$\avr B F = f(F).$$

The correlation function can be approximated by the linear function: $\avr B F = a+b_{BF} F$ using the linear regression.
Here $b_{BF}$ is the correlation coefficient, a slope of correlation function

Alternative definition of the correlation coefficient is
\beq\label{bFB}
b_{BF}=\frac{\av{FB}-\av{F}\av{B}}{\av{F^2}-\av{F}^2}=\frac{\mathrm{cov}(F,B)}{D_F}.
\eeq

For a correlation between relative variables,  $F/\av F$ and $B/\av B$:
\beq\label{brelFB}
\brel_{FB} = \frac{\av F}{\av B} b_{FB} \ .
\eeq

For the observables, $B$ and $F$, in backward and
 forward rapidity windows,  multiplicity ($n$)
 or mean event transverse momentum ($\text{p}_\text{t}$=$\frac{1}{n}\sum\limits_{i=1}^{n} {\text{p}_\text{t}}_i$) can be taken \cite{Braun:2003fn,Vechernin:2003vcen,Vechernin:2003wpen}.
 
Note that the multiplicities in backward and forward rapidity windows $n_B$, $n_F$ are extensive variables, which means that they are proportional to the size of the system. In contrast, the transverse momentum of particles in backward and forward rapidity windows  $p_{tB}$,  $p_{tF}$ are intensive variables, so they don't depend on the number of sources which produce particles.

The long-range multiplicity correlation $b_{nn}$ at large rapidity gap is dominated by event-by-event variance in the number of sources (cut pomerons, strings)\cite{Capella:1977me}. In particular, in nuclear collisions, it also reflects the  fluctuation in the number of participant nucleons. The influence of possible string fusion processes on this type of correlation is rather small \cite{Braun:1999hv}.  It also has been shown, that $b_{nn}$ depend strongly on the method of the centrality determination and the on the width of centrality class \cite{Kovalenko:2013jya,Sputowska:2223886,Drozhzhova:2016njd}.

However, the event-by-event fluctuations in the number of cut pomerons (strings)
(the ``volume'' fluctuations)  do not lead to the correlation
between the intensive variables, e.g. the $p_{tB}$-$p_{tF}$ correlation ($b_{p_t\text{-}p_t}$). 
This makes ${\text{p}_\text{t}\text{-}\text{p}_\text{t}}$ correlations robust against the volume fluctuations and the details of the centrality selection.
Rather the long-range $p_{tB}$-$p_{tF}$ correlation  indicates the fluctuations in ``quality'' of sources.




\section{String Fusion Model}

Due to the fact that quark-gluon strings have the finite transverse size they can overlap in the impact parameter plane. 
The interaction of color strings can be performed in the framework of local string fusion model \cite{Braun:1999hv}. 

According to this model, mean multiplicity of charged particles and mean $p_t$ originated from the area $S_k$, where k strings are overlapping are the following:
\begin{equation} \label{muptloc}\nonumber
	\left\langle \mu\right\rangle_k=\mu_0 \sqrt{k} \frac{S_k}{\sigma_0}, \hspace*{1cm}
	\left\langle p_t\right\rangle_k=p_0 \sqrt[4]{k},
\end{equation}	 
where $\sigma_0=\pi r_{str}^2$ -- string transverse area.

The calculations are simplified in a cellular variant of the string fusion model by the introduction of the lattice in the impact parameter plane \cite{Vechernin:2007zz, Vechernin:2007zza, Braun:2003fn, Kovalenko:2012ye,Vechernin:2003wpen}. In the discrete model, the transverse plane is considered as a grid with the cell area equals to string transverse area, and strings are fused if their centers occupy the same cell.
The cellular variant of string fusion has been shown to provide the results
very close to the ones in the original string fusion model \cite{Lakomov:2013jna}.

\section{Dipole-based MC SFM model}
The Monte Carlo model \cite{Kovalenko:2012ye,Kovalenko:2012nt, Kovalenko:2013jya} used in this paper is based on dipole picture of elementary partonic interactions. The model preserves energy and angular momentum conservation in the initial state of a nucleon \cite{Kovalenko:2012nt,Kovalenko:2013oua}.
The probability of dipoles interaction depends on their
transverse coordinates $\vec{r}_1,\vec{r}_2,\vec{r}_1',\vec{r}_2'$ with effective coupling:
\begin{equation}\label{wlog}
f=\dfrac{\alpha_s^2}{2} \ln^2 \dfrac{|\vec{r}_1-\vec{r}_1'||\vec{r}_2-\vec{r}_2'|}{|\vec{r}_1-\vec{r}_2'||\vec{r}_2-\vec{r}_1'|},
\end{equation}
where $(\vec{r}_1, \vec{r}_2), (\vec{r}_1', \vec{r}_2')$ are transverse coordinates of the projectile and target dipoles. \newline
 With confinement effects taking into account, 
\cite{Gustafson:2009qz, Flensburg:2011kk}, 
the probability amplitude is:
\begin{equation} \label{newformula}
			f=\frac{\alpha_S^2}{2}\Big[ K_0\left(\frac{|\vec{r}_1-\vec{r}_1'|}{r_{\text{max}}}\right) +
			K_0\left(\frac{|\vec{r}_2-\vec{r}_2'|}{r_{\text{max}}}\right) 
			- K_0\left(\frac{|\vec{r}_1-\vec{r}_2'|}{r_{\text{max}}}\right)
			- K_0\left(\frac{|\vec{r}_2-\vec{r}_1'|}{r_{\text{max}}}\right)	\Big]^2.
\end{equation}	
Here $r_{max}$ is characteristic confinement scale.

Multiplicity and transverse momentum in the model are calculated in the approach of color strings, stretched between projectile and target partons, taking into account their finite rapidity width and local cellular string fusion.
Strings are assumed to emit particles independently according to Poisson distribution (no short-range effects are included).
Parameters of the model are constrained from the p-p, p-Pb and Pb-Pb data on total inelastic cross section and multiplicity \cite{Kovalenko:2014tca}.

\section{Results in dipole-based MC SFM model}
\begin{figure}[h]
\begin{center}
{\resizebox{0.48\columnwidth}{!}
{\includegraphics{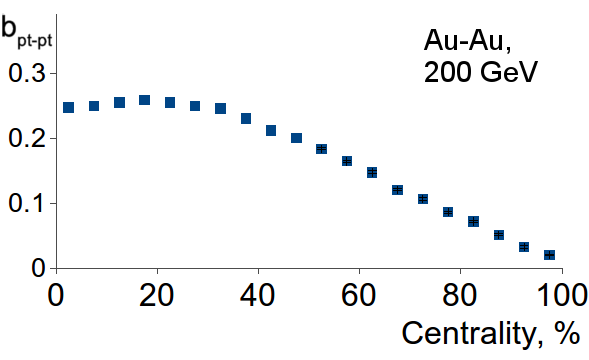}}}
{\resizebox{0.50\columnwidth}{!}
{\includegraphics{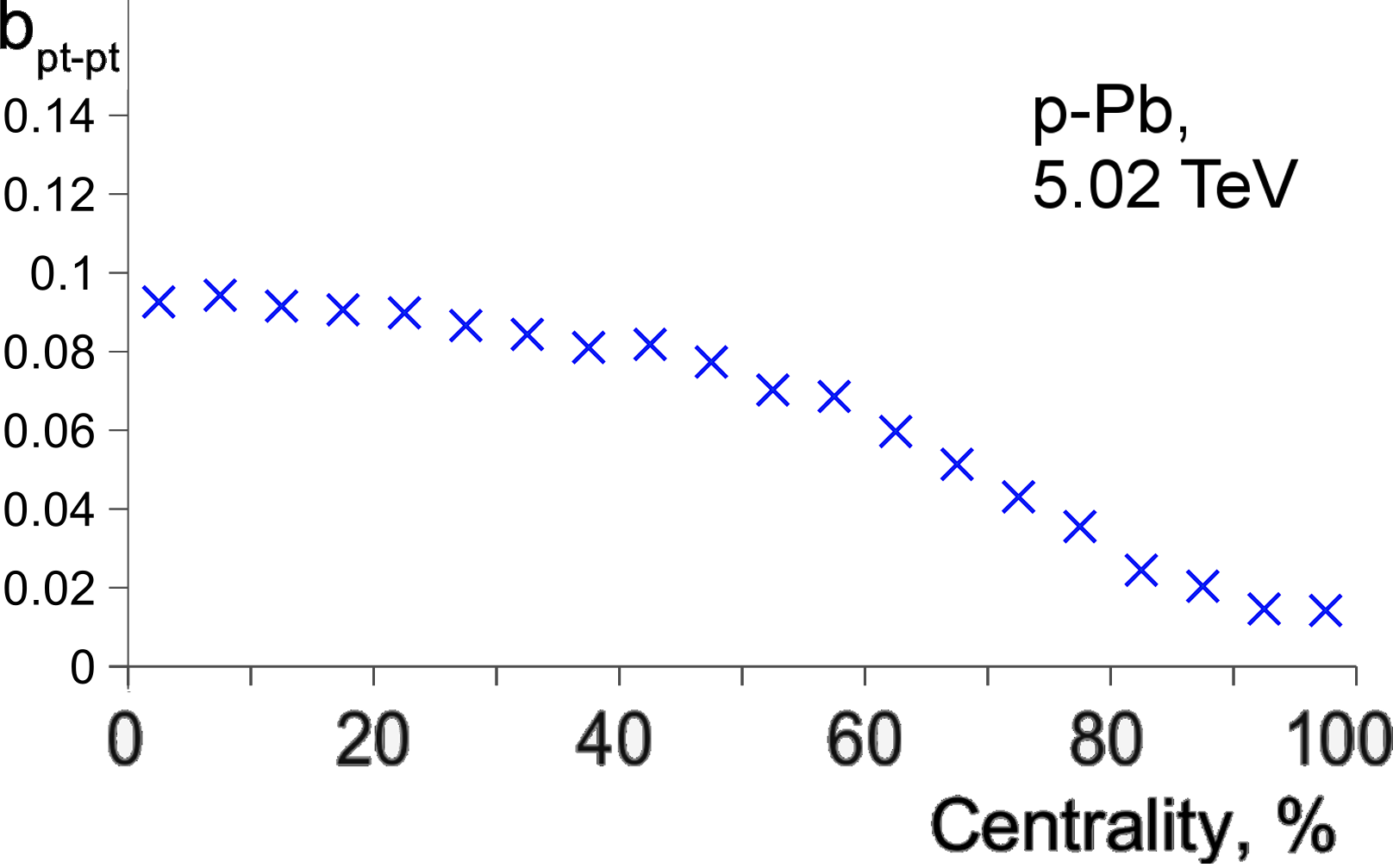}}}
{\resizebox{0.49\columnwidth}{!}
{\includegraphics{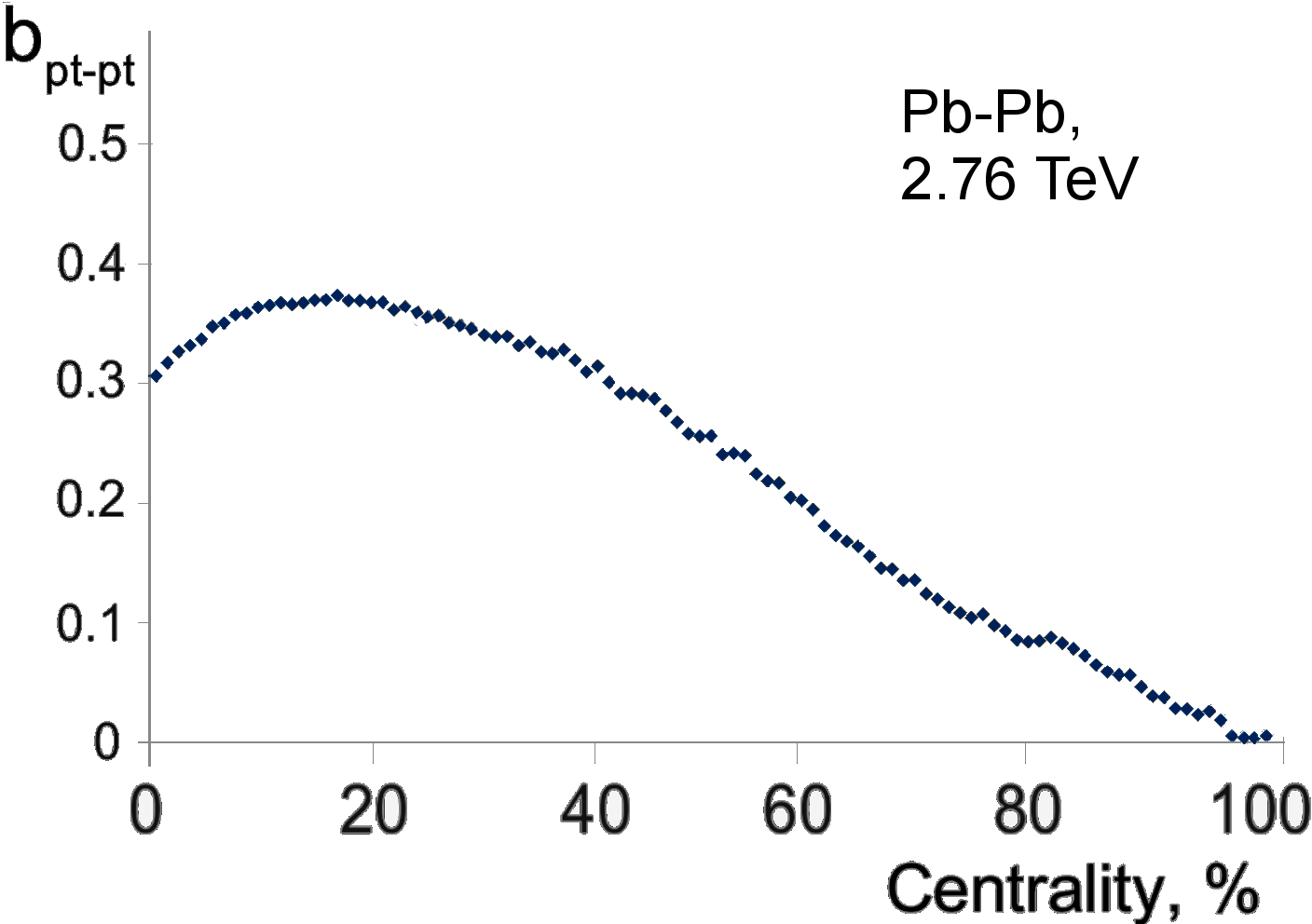}}}
 \end{center}
\caption{ 
Centrality dependence of $b_{p_t-p_t}$ for  Au-Au collisinos at $\sqrt{s_{NN}}$=200 GeV (top~left), $\text{{p-Pb}}$ collisinos at $\sqrt{s_{NN}}$=5.02 TeV (top right) and Pb-Pb collisions at $\sqrt{s_{NN}}$=2.76 TeV (bottom). MC simulations at $r_\mathrm{str}$=0.2~fm.}\label{fig:figBptptPbPb}
\end{figure}

In dipole-based Monte Carlo string fusion model, 
we calculated the mean transverse momentum correlation coefficient ($b_{pt-pt}$) for Au-Au collisions RHIC energy and p-Pb and Pb-Pb collisions at the LHC energies.
The results are shown in figure \ref{fig:figBptptPbPb}.

The results show a general growth of $b_{p_t\text{-}p_t}$ with centrality reflecting the increase of string density and, hence, the role of
string fusion from peripheral to central collisions.
We note that in Pb-Pb collisions at the LHC energy the correlation coefficient 
has a maximum in mid-central collisions and further decreases with 
centrality. This regime reflects the attenuation of color field
fluctuations due to the string fusion at large string density.

We note that the non-monotonic behavior of $b_{pt-pt}$ with centrality is achieved only in heavy ion collisions at the LHC, while in Au-Au collisions at RHIC and p-Pb at LHC the maximal string density is not enough to provide a decline of the correlation coefficient for most
central collisions.

\section{Mean ${\text{p}_\text{t}\text{-}\text{p}_\text{t}}$ correlations in different models}

We studied the mean ${\text{p}_\text{t}\text{-}\text{p}_\text{t}}$ correlations LHC energy in different Monte Carlo generators. The following Monte Carlo model were considered:

\begin{itemize}
\item MC model with  string fusion \cite{Kovalenko:2012nt,Kovalenko:2013jya} (see above).
\item { THERMINATOR 2} (THERMal heavy IoN generATOR 2)
\cite{Chojnacki:2011hb}. Based on
parametrized freeze-out hypersurface, Cooper-Frye particlization and decays.
\item { HIJING} (Heavy Ion Jet INteraction Generator)
\cite{Wang:1991hta}. It includes  gluon shadowing effects and jet quenching.
\item { DPMJET}, two-component Dual Parton Model, based on the Gribov-Glauber approach
\cite{Roesler:2000he}.
It models soft and hard components and performs fragmentation of partons by the Lund model.
\item { AMPT} (A Multi-Phase Transport Model for Relativistic Heavy Ion Collisions)
\cite{Lin:2004en}.
It includes gluon shadowing, Zhang's Parton Cascade, string melting,  relativistic transport.
\end{itemize}

\begin{figure}[h] \vspace{0.4cm}
\begin{center}
\begin{overpic}[width=0.99\textwidth]{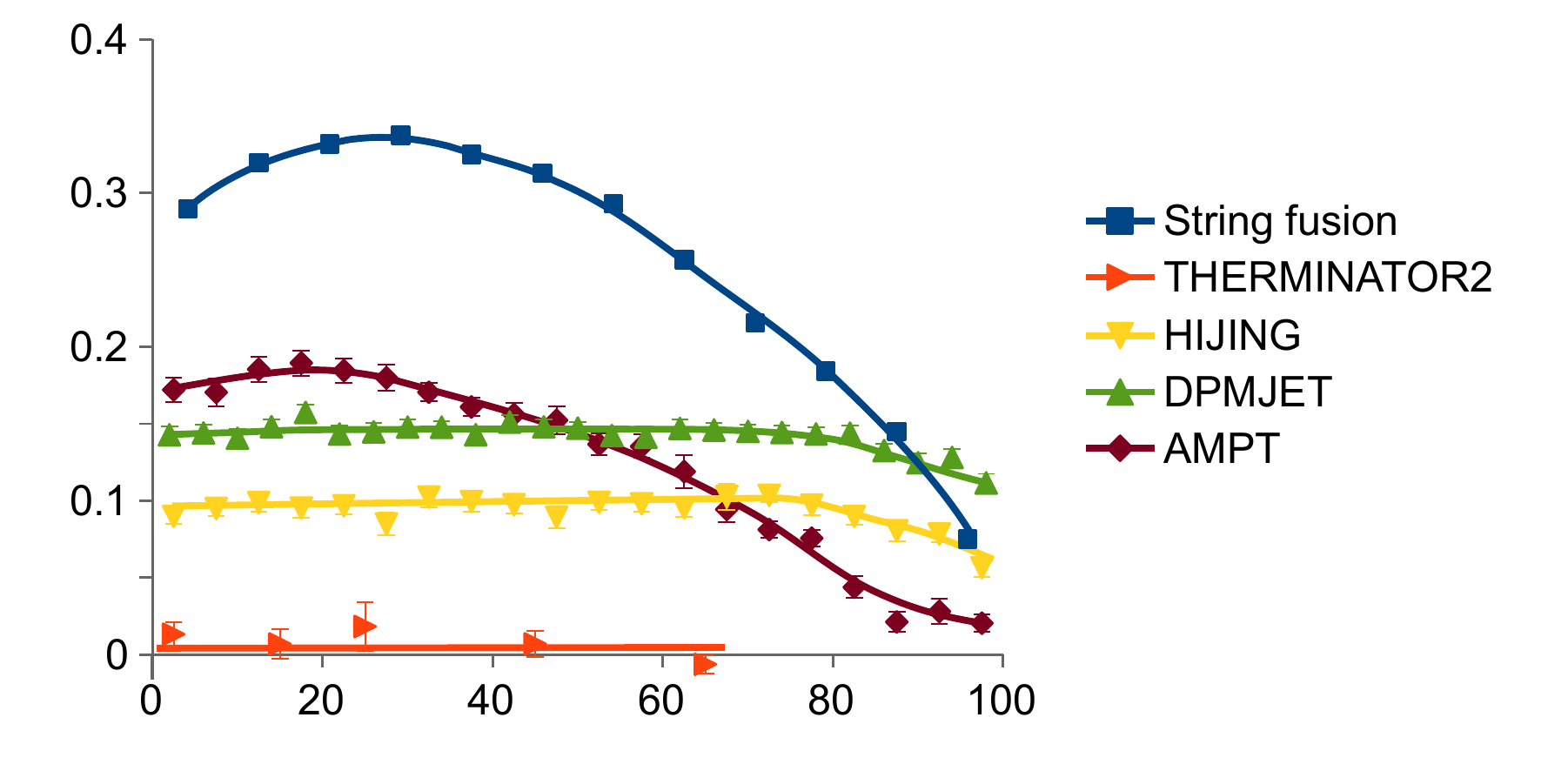}
 \put (4,50) {\large$\displaystyle b_{p_t\text{-}p_t}$}
\end{overpic}
 \end{center}\vspace{-0.4cm}
\caption{ \label{fig:figall_s_5}
Centrality dependence of $b_{p_t\text{-}p_t}$ for Pb-Pb collisions at $\sqrt{s_{NN}}$=2.76 TeV in different Monte Carlo models. }

\end{figure}

The results for $\text{p}_\text{t}\text{-}\text{p}_\text{t}$
correlations in Pb-Pb collisions at LHC energy are shown in figure \ref{fig:figall_s_5}.

The results demonstrate that the model with parameterized initial states shows zero correlation coefficient. In the models, which include only initial state shadowing and include soft+hard components, the forward-backward ${\text{p}_\text{t}\text{-}\text{p}_\text{t}}$ correlation is small and independent on centrality.

The relativistic transport model and string fusion model
demonstrate significant non-trivial centrality dependence of $b_{p_t\text{-}p_t}$ as well as its non-monotonic behavior.

The comparison clearly shows that ${\text{p}_\text{t}\text{-}\text{p}_\text{t}}$ forward-backward correlation and its centrality dependence is  sensitive to the initial stages of 
heavy ion collisions.

\section{Summary and conclusions}
The dependence of the correlation strength between mean-event transverse momenta on the collision centrality is obtained for different collision energies. It is shown that above RHIC energy the dependence reveals the decline of the correlation coefficient for most central collisions, reflecting the attenuation of color field fluctuations due to the string fusion at large string density.



The long-range correlation between intensive observables, being robust against the volume fluctuations and the details of the centrality determination, enables to obtain the signatures of string fusion at the initial stage of hadronic interaction in relativistic heavy ion collisions at LHC energy.

%
The sensitivity of the long-range ${\text{p}_\text{t}\text{-}\text{p}_\text{t}}$ correlations to 
the properties of the initial stages of heavy ion collisions
can be further studied in fully event-by-event
hydrodynamical models, such as iEBE-VISHNU \cite{Shen:2014vra} or EKRT \cite{Niemi:2015qia}.

The authors acknowledge Saint-Petersburg State University for the research grant 11.38.242.2015.

\section*{References}
\bibliography{bibtex.bib}	

\end{document}